\documentclass[conference]{IEEEtran}
\usepackage[colorlinks,citecolor=blue,urlcolor=blue,bookmarks=false,hypertexnames=true]{hyperref}
\usepackage[utf8]{inputenc}
\usepackage[T1]{fontenc}

\usepackage[ngerman, american]{babel} 
\usepackage[final,babel]{microtype}

\usepackage[normalem]{ulem}

\usepackage{pifont}

\usepackage[caption=false,font=footnotesize]{subfig}
\usepackage[
  backend=biber,
  style=ieee,
  minnames=1,
  maxcitenames=2, maxbibnames=2,
  url=false, doi=false, 
  isbn=false, eprint=false, 
]{biblatex}

\makeatletter
\def\blx@citeadd#1{%
	\ifcsdef{blx@keyalias@\the\c@refsection @#1}
	{\edef\blx@realkey{\csuse{blx@keyalias@\the\c@refsection @#1}}}
	{\def\blx@realkey{#1}}%
	\expandafter\blx@getrefcontext\expandafter{\blx@realkey}
	\expandafter\blx@citation\expandafter{\blx@realkey}\blx@msg@cundefon
	\expandafter\blx@ifdata\expandafter{\blx@realkey}
	{\advance\blx@tempcnta\@ne
		\listeadd\blx@tempa{\blx@realkey}}
	{\ifnum\blx@tempcntb>\z@\multicitedelim\fi
		\expandafter\abx@missing\expandafter{?}
		\advance\blx@tempcntb\@ne}}
\makeatother

\usepackage[dvipsnames,table]{xcolor}

\usepackage[english,ruled,vlined,linesnumbered,algosection,algo2e]{algorithm2e}

\usepackage{siunitx}
\usepackage{xfrac}
\ifCLASSINFOpdf
\else
\fi
\usepackage{graphicx}
\usepackage{placeins}
\usepackage{subcaption}
\usepackage{adjustbox}

\usepackage{tikz}

\usepackage{longtable}
\usepackage{booktabs}
\usepackage{multirow}
\usepackage{multicol}
\usepackage{colortbl}

\usepackage{eqparbox}

\usepackage{textcomp,marvosym}
\usepackage{latexsym,amssymb, amsfonts}
\usepackage[intlimits]{amsmath}

\usepackage{fixmath, upgreek}

\usepackage{blindtext}

\renewcommand{\texttt}[1]{{\small \ttfamily #1}}

\usepackage{sourcecodepro}
\makeatletter 
\def\verbatim@font{\fontfamily{SourceCodePro-TOsF}\footnotesize
\let\do\do@noligs 
\verbatim@nolig@list} 
\makeatother

\usepackage[macros=true, xspace=true]{acro}
\DeclareAcronym{TRNG} {short = TRNG,long = True Random Number Generator,tag = abbrev}
\DeclareAcronym{STR} {short = STR,long = Self-timed Ring,tag = abbrev}
\DeclareAcronym{FPGA} {short = FPGA,long = Field Programmable Gate Array,tag = abbrev}
\DeclareAcronym{DUT} {short = DUT,long = Device Under Test,tag = abbrev}

\begin{document}

\title{Side-Channel Analysis of the RP2350 Challenge 2}

\author{\IEEEauthorblockN{Nikhil Tattikota}
\IEEEauthorblockA{School of Electrical Engineering and Computer Science\\
Oregon State University\\
Corvallis, OR, USA\\
Email: tattikon@oregonstate.edu}
\and
\IEEEauthorblockN{Vincent Immler}
\IEEEauthorblockA{School of Electrical Engineering and Computer Science\\
Oregon State University\\
Corvallis, OR, USA\\
Email: immlerv@oregonstate.edu}
}

\maketitle

\begin{abstract}
Bug bounty programs and community challenges have become an established mechanism to incentivize collective action to improve security.
In essence, these can be considered as crowd-sourcing mechanisms where participants carry the risk twice over: performing the work in addition to having it officially accepted, which at times can be tied to unacceptable terms.
Consequently, independent assessments and publications that are free from third party obligations remain important to highlight deficiencies in implementations of public significance.

We consider the RP2350 side-channel analysis challenge and its hardened Advanced Encryption Standard (AES) implementation to be of reasonable importance to justify such an approach, especially so since the presented challenge remained unsolved for more than a year.
In our paper, we present our analysis, demonstrating comprehensive leakage detection, plaintext dependence, and key extraction from the challenge implementation.

Our work presents concerning weaknesses in the implementation that contradict the challenge authors' claims.
We hope by openly sharing this analysis to motivate further public scrutiny.
\end{abstract}


\IEEEpeerreviewmaketitle

\section{Introduction}

Secure communication and data storage for a wide range of purposes relies on electronic systems that guarantee integrity and confidentiality of their data.
Mainly, two different categories of products exist that address this: \emph{i}) standard products with limited range of hardware cryptographic functions that are not security-certified, such as the RP2350 microcontroller, and \emph{ii}) products with countermeasures where security was assessed by accredited test labs, to obtain certifications such as Common Criteria (CC)~\cite{CCv3.1}. 
Based on the specifics of the certification, the finesse and scope of countermeasures can vary greatly. 

Especially non-invasive attacks such as Differential Power Analysis (DPA)~\cite{MOP07} and Electro-Magnetic Analysis (EMA) are considered a substantial threat, as the attacker's resources can be quite limited compared to more advanced and expensive attacks.
This is especially true since the ChipWhisperer~\cite{OC14} was popularized for power analysis, a single-channel oscilloscope with built-in clock generator at a price-performance point where equally capable oscilloscopes are typically more expensive.

Achieving a secure implementation requires a robust development and subsequent security evaluation of the countermeasures based on leakage detection and attempted key extraction.
Typically, a wide range of measurements must be performed under varying physical conditions such as type of measurement, Region of Interest (ROI) on a chip, equipment used, and different logical inputs (data acquisition patterns) as stimuli.
This often result in vast amounts of data with substantial differences in the assessed security levels of different data sets, despite observing the same implementation~\cite{MM12,ISU18,SIU+18,BUS22}.

To counteract this, the SCA community has been exploring standardized side-channel comparison platforms to essentially enable `best case' test scenarios.
However, individual designs reflect engineering decisions by their author(s), leading to inconsistent results even across competing designs of such boards~\cite{FBAK25}.
Consequently, it remains imperative to tailor the analysis approach to the Device Under Test (DUT) and not exclusively rely on standardized ecosystems that can be limited.

To address this need, the Raspberry Pi foundation organized the RP2350 hacking challenge 2 (RP2350hc2)~\cite{RaspberryPi26}.
Challenge participants are asked to investigate the side-channel security of an AES256 implementation.
Participants are encouraged to reach out to the organizers by stating: ``\emph{If you think you're onto something [...], we'd like to hear from you early}'' while pointing out that it is a ``\emph{first across the line}'' competition with a single reward, raising concerns as to how valuable work from other participants will be honored, which may explain why the challenge has not garnered the desired attention, resulting in a \emph{three}-times extended deadline. 
In the following, we present our independent findings for this hacking challenge.

\subsection{Motivation}

A hardening report is provided by the challenge authors that describes their attempts to secure the AES implementation, in addition to the source code itself.
Our work is motivated by two cues in the report that are worth exploring:
\begin{itemize}
\item It appears, only a ChipWhisperer Husky (Plus) was used for leakage detection, i.e., no lab-grade instruments
\item Absence of a formal security level in combination with no mention of higher-order \emph{moment} attacks; only of multivariate higher-order attacks (combination of points)
\end{itemize}

The first point is of interest as the ChipWhisperer Husky (Plus) as a product does \emph{not} have a specified bandwidth!
Its internal AD8330 Variable Gain Amplifier (VGA) has a $-3\,\mathrm{dB}$ bandwidth of $150\,\mathrm{MHz}$, i.e., even under ideal conditions, the fundamental frequency of the RP2350 clock of $150\,\mathrm{MHz}$ is only observed with a $-3\,\mathrm{dB}$ loss already.

However, its other limit is even more relevant: a sample rate of 250 MS/s (Husky Plus), resulting in a Nyquist frequency of only $125\,\mathrm{MHz}$ that can be accurately recorded.
This contradicts best practices for data collection that require a \emph{minimum} sample rate of 5x the bandwidth, with an input bandwidth close to the device clock frequency~\cite{GJJR11}.
In addition, the RP2350 relies on a Phase-Locked-Loop (PLL) which, even when synchronizing through its clock input, results only in a quasi-synchronous setup where additional timing uncertainty is easily created by an improper choice of shunt resistor.
Consequently, the ChipWhisperer is not a suitable evaluation tool for this target.

The second point is of interest, as higher-order attacks can be considered a combination of points (multivariate analysis), as guessing additional (mask) values, and/or analysis of higher-order moments.
Since their report does not mention higher-order moments, and no formal security argument is provided, it is unclear what the designed-for security level is and if corresponding analysis techniques have been explored.

Please note that targeting the $n$-th statistical moment results in an exponential increase in data complexity which is why a low-noise setup is important~\cite{SM15,MM17}.
Consequently, as a result of apparent limitations in their measurement setup, combined with weak(er) leakage detection as referenced in their report, a less-than-optimal hardening is the expected result. 

\subsection{Contribution}
In this paper, we present the following contributions:

\begin{itemize}
    \item (\emph{Unendorsed}) Results of the RP2350 challenge 2, presenting a side-channel evaluation based on leakage detection, plaintext dependence, and key extraction
    \item Analysis of selected practical aspects that are relevant to the success of performing a side-channel analysis, especially within the presence of countermeasures
    \item Important discussion of community challenges to encourage meaningful improvements for future challenges
\end{itemize}

\subsection{Discussion of the Challenge (Participant Perspective)}

The challenge was published July 29, 2025.
After initially going unsolved, certain ``hiding'' countermeasures were removed on February 3, 2026.
The challenge was ultimately extended to Oct 31, 2026.
The RP2350 hacking challenge 2 GitHub states the following two conditions to win:

\begin{enumerate}
\item ``\emph{[\dots] to leak AES key material in a way that will reduce the effective key length [\dots]}''
\item ``\emph{A winning method must also be possible using the actual self-decrytping binary/bootloader}''
\end{enumerate}

We consider the second condition to be underspecified, as no separate firmware other than the challenge firmware was provided.
No attack limitations were imposed that would substantially explain the difference between 1) and 2).
In an attempt to clarify and engage with them, we have been in touch with the RP2350 challenge organizers as follows:

\begin{itemize}
\item Inquiry about the exact criteria to win on July 1, 2026
\item Starting our work with leakage detection in August 2026
\item Receiving additional criteria to win the challenge on September 2, 2026; confirmation by the challenge organizers that no winning submission had been received.
\item Reporting several more results until the challenge concluded (proposed attack) and beyond; including this paper
\end{itemize}

These additional criteria were not publicly disclosed on their GitHub page and would have involved providing \emph{all} of our internal software (oscilloscope automation, data stimulus framework for target, analysis scripts, etc.).
The paper authors deemed these criteria as unacceptable, as no licensing would control how they could distribute our software.
Additional reasons include: ethical concerns over providing a turnkey solution that would enable extraction as a service, legal concerns over export controls of dual-use software, fairness concerns towards other contestants using commercial software, and a reward that is already below typical industry rates for the analysis work or even just licensing costs of commercial side-channel analysis software (which we do not have available).

Since the organizers emphasized that only a single winner would be selected, it is surprising as to why they let the challenge deadline expire only then to confirm having received multiple submissions, instead of closing the challenge upon having received the first submission.
It is undeniably beneficial to the organizers to let authors continue submit their findings despite having received a submission already. 
We appeal to the challenge organizers to develop an improved framework for future challenges to value all reasonable contributions.

We acknowledge the need of the organizers to verify submissions from participants.
Accordingly, the challenge organizers were offered to inspect our internal software, acquisition process, and analysis in a read-only environment to ensure that no inadvertent mistakes happened that could have weakened the security of the implementation.
We also kindly offered our measurement data to them under the condition they provide us with the necessary physical storage.

Except minor modifications of their firmware to make measurements more convenient (e.g., increasing the UART speed), we want to emphasize that the AES itself and the trigger were \emph{not} modified in any way.
In addition, any other relevant setting of the microcontroller, such as clock speed or status of peripherals, was left unchanged.

\subsection{Paper Outline}

The remainder of the paper is organized as follows. 
In Section~\ref{sec:relatedwork}, we discuss related work on side-channel analysis and provide relevant background information, in particular on side-channel analysis countermeasures.
Afterwards the scope of our analysis and our measurement setup are explained in Section~\ref{sec:scope_and_setup}.
Subsequently, in Section~\ref{sec:impldetails}, we provide a high-level assessment of the challenge implementation.
This leads to the leakage detection and key extraction results as presented in Section~\ref{sec:sca_analysis}.
Our conclusions are provided in Section~\ref{sec:concl}.

\section{Related Work and Background}\label{sec:relatedwork}

\subsection{Side-Channel Analysis}

This work is in line with previous evaluations that were published within an academic context.
Two of these works \cite{MH15, CEM18} also discuss relevant low-power measurement techniques.

Since the challenge is concerned with real-world attacks, we strictly focus on techniques not requiring profiling~\cite{CRR03}.
In other words: a reasonably motivated attacker could perform this analysis in a black-box setting, emphasizing the relevance of our results.
In addition, key updates turned out to be too slow on the target (unless modified appropriately), which further encouraged us to only consider non-profiled techniques.

Typical steps of a side-channel analysis are leakage detection~\cite{GJJR11,SM15,MWM21} and attempted key extraction~\cite{MOP07}.
Over the past couple of years, many analysis-oriented works have focused on deep-learning~\cite{PPM+23}.
However, these techniques often cannot be performed in a non-profiled way.

A welcome exception is SCATTER~\cite{TGWC17} that is rooted in Mutual Information Analysis~\cite{GBTP08}.
SCATTER creates trace distributions based on a sliding window to transform temporal trace portions into untemporal distributions.
This can be performed in a univariate-style scenario, but also in a higher-order fashion where different ranges are combined.
Especially in the presence of hiding countermeasures~\cite{MOP07} such as jitter, or permutation shuffles, higher-order SCATTER combined with Spatial Dependency Analysis (SDA) as presented in~\cite{VTM21} is a powerful tool to evaluate complex side-channel mixtures.

This work discusses the application of these techniques in the presence of countermeasures.
In particular, we discuss the unique benefits of combining some of these analysis techniques.

\subsection{Side-Channel Countermeasures}

Different countermeasures~\cite{MOP07} in response to DPA attacks have been published, often with a particular focus on either software~\cite{CBCH23} or hardware~\cite{CEM18}.
So called ``hiding'' countermeasures try to reduce data dependency in the power consumption, e.g.,  by randomizing the sequence of operations or using tailored logic styles to enforce a data independent switching activity.
Other countermeasures attempt to break the dependency between input values and sensitive intermediate values by ``masking'' at the algorithmic level.
Real-world implementations typically use a combination of several countermeasures.

For masking, sophisticated concepts exist such as Threshold Implementations~(TIs)~\cite{CEM18}.
They split the sensitive intermediate value into multiple independent shares.
To achieve a formal level of security, $d$th-order secure TIs require at least $d+1$ shares that must fulfill three properties -- correctness, non-completeness, and uniformity.
Non-completeness is achieved by implementing a set of tailored functions, where each works on a subset of the shares only. 
The output of these functions are then combined to produce the designated result, thereby fulfilling the correctness property.
The third, and most challenging property to achieve is uniformity of the tailored output functions.

Other masking schemes exist that in part, are rooted in the same ideas.
Table-based masking~\cite{RDP08} equally splits the values into shares, however, it uses pre-computed split-lookup tables to handle non-linear steps of the algorithm.
These schemes additionally need to take into account potential leakage of intermediate indices and prevent transitional leaks by explicitly guarding against them.

When implementing these countermeasures, different practical aspects and leakage models must be considered~\cite{CEM18,CBCH23}.
For example, a $d$th-order secure scheme in a value-based leakage model, will only provide $\lfloor \frac{d}{2} \rfloor$ order in the transition-based leakage model based on the security-order reduction theorem~\cite{BGG+15}.
Additional aspects must be considered for both implementation and measurement.
Coupling effects, (un)intentional leakage overlap, and several other factors can quickly result in diminished security~\cite{CEM18,MM13,CBCH23}.

However, even when ignoring these implementation challenges, masked implementations can be broken by $d\,+\,1$-th-order side channel attacks.
This can be either achieved by a combination of the right points or by estimating higher-order statistical moments.
Alternatively, leakage can also be converted from multivariate to univariate~\cite{MM13}, or from higher-order moment to lower-order moment~\cite{MM17}.

\section{Scope of Analysis and Setup} \label{sec:scope_and_setup}

\subsection{Scope of Anlaysis}

Our work was time-limited, focusing on \emph{i}) Solving the challenge task in a black-box style setting, and \emph{ii}) Develop a rationale as to why the attack works.
Consequently, we consider the following aspects to be out of scope for this paper: 
source code, while available, was only studied to the extent of confirming basic side-channel observations; 
we did not perform an analysis of the provided emulator or virtual power-traces; 
no attempts were made to improve their implementation.

\subsection{Measurement Setup for the Analysis} \label{sec:measurementsetup}

\paragraph{Device Under Test (DUT)}

\begin{figure}[!tbh]
    \centering
    \subfloat[Top of modified RP2350\label{fig:commercial_probe_draw_subfig}]{
        \includegraphics[width=0.4\columnwidth]{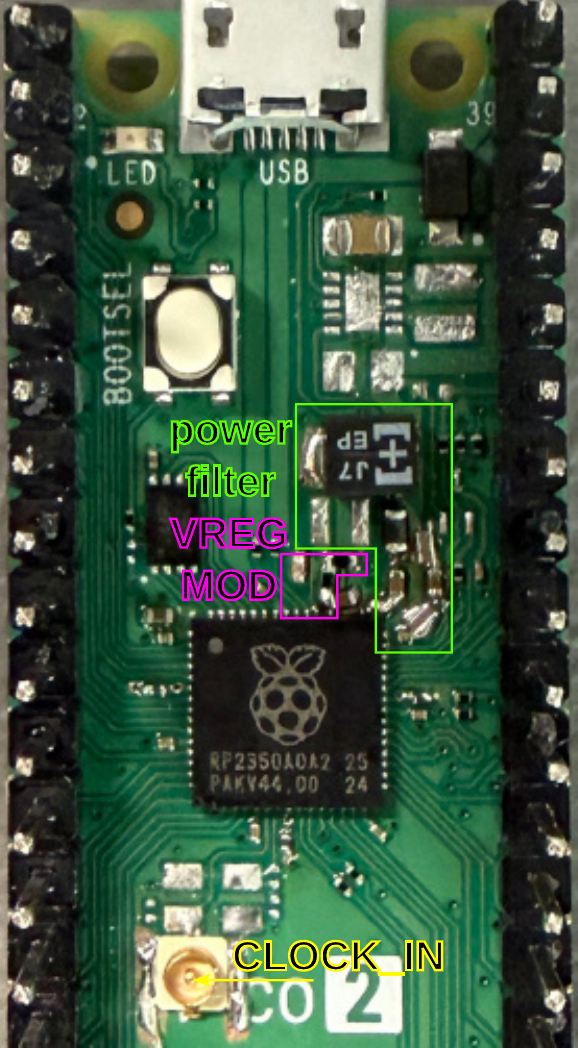}
    }
    \subfloat[Bottom of modified RP2350\label{fig:oblique_probe_draw_subfig}]{
        \includegraphics[width=0.41\columnwidth]{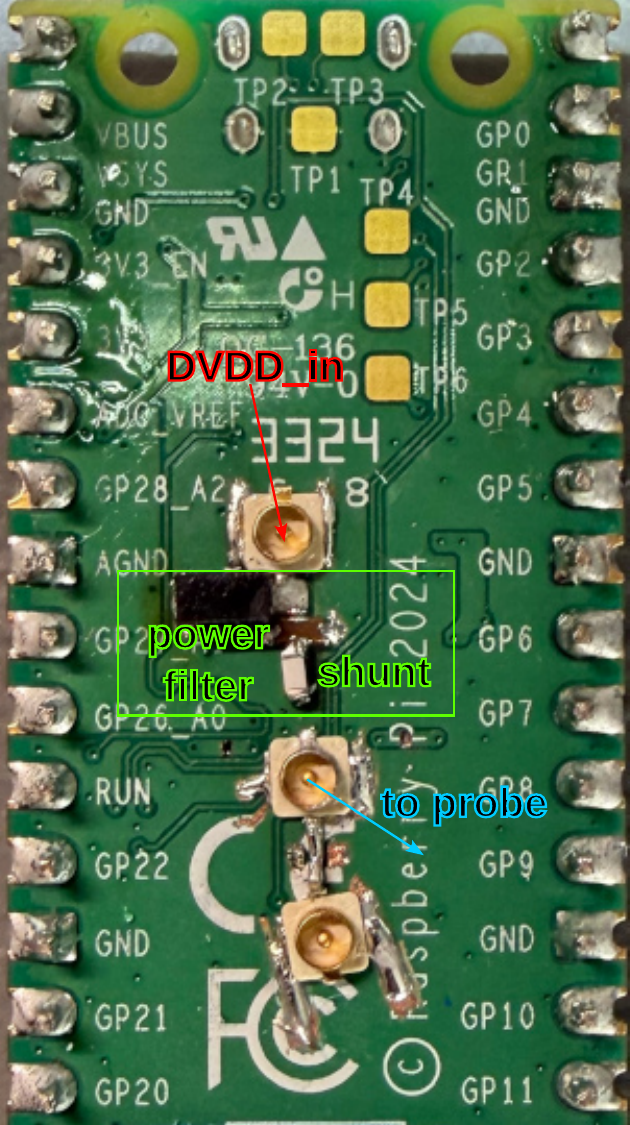}
    }
    \caption{``Octopus''-style modified RP2350 DUT.}
    \label{fig:rp2350mods}
\end{figure}

Our DUT is a standard RP2350 development board with A2 revision of the chip (cf. Figure~\ref{fig:rp2350mods}). 
Different PCB revisions exist with a differently sized DVDD decoupling capacitor on one side of the chip, including minor layout modifications.
We are unaware of any impact of the board revisions on the analysis success.
Proper preparation of the board proved to be highly relevant to success.

Partially outdated side-channel analysis wisdom is to ``\emph{place the highest value [shunt] resistor [\dots] that allows the device to function.}''~\cite{GJJR11}. 
However, the RP2350 PLL is very sensitive to the shunt resistor value in the DVDD rail. 
Too large values result in strong PLL sideband spurs, sometimes even lopsided.
Based on the load profile, local PLL instability can be observed. 

A more balanced result is achieved with resistor values in the range of (only) up to $1 \,\Omega$, or an inductive shunt, and an increased DVDD voltage of $1.3-1.5\,\mathrm{V}$.
These findings align with~\cite{CEM18}, where lower-order moment leakage was easier to observe for otherwise higher-order protected implementations when the voltage was increased, and the shunt \emph{removed}.
Our board modifications are summarized as follows and illustrated in Figure~\ref{fig:rp2350mods}: \emph{i}) removal of all factory-provided DVDD capacitors, \emph{ii}) removal of oscillator to supply external clock, \emph{iii}) set fixed reference for VREG\_FB using resistor ladder, \emph{iv}) in-situ PCB modifications to supply power from two external LT3045 regulators, \emph{v}) modifications for additional power filtering.
Our results demonstrate that the attack can be successfully performed on a commercial off-the-shelf RP2350 board.

\paragraph{Oscilloscope}

\begin{figure}
\centering
\includegraphics[width=0.5\columnwidth]{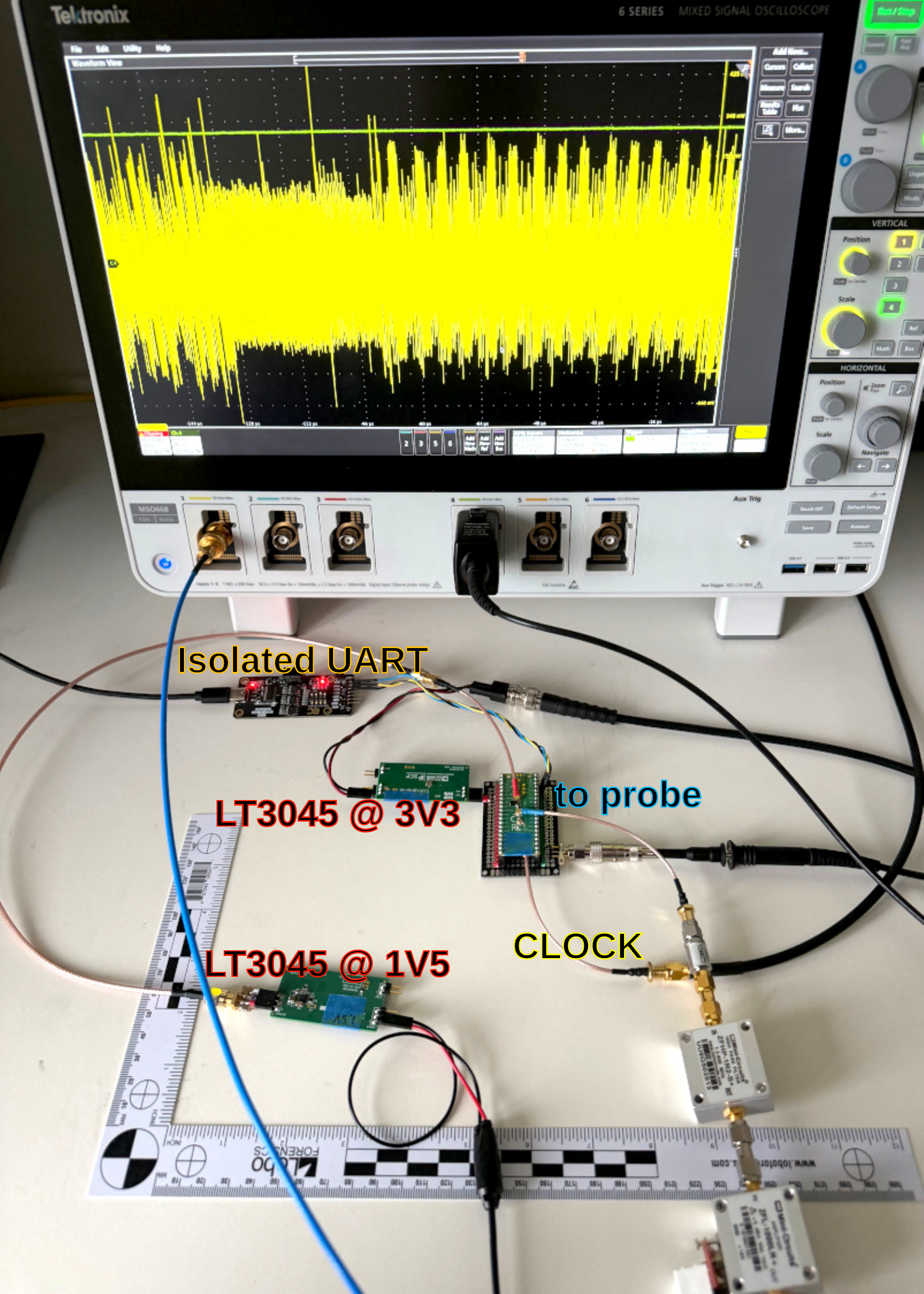}
\caption{Measurement setup for RP2350 DUT.}
\label{fig:setup}
\end{figure}

We use a Tektronix MSO66B (cf. Figure~\ref{fig:setup}) at a sample rate of 625 MS/s and a bandwidth limit of 200~MHz.
The DUT was clock-synchronized to the oscilloscope's internal signal generator at 12 MHz.

The RP2350hc2 AES provides a trigger that is high during the AES computation.
We trigger on the falling edge and use pre-trigger samples only.
A full recording of the challenge firmware is 300 kpts.
However, only the final 100 kpts need to be recorded (and far fewer points stored) for the attack to work.
Oscilloscope automation was done with TekHSI, resulting in approx. 59 traces per second, resulting in a practical upper limit of about 5 M to 10 M traces (1 or 2 days) for the analysis.

\paragraph{Probing}

The target exhibits a low power behavior where power ripple at 1X can be observed with an oscilloscope at approx. 3-4 mV/div.
Side-channel dependent variations are even smaller, suggesting moderate signal amplification. 

We use a single-ended voltage measurement based on a DC block and a ZFL-1000LN+ amplifier. 
An additional ZFHP-1R2-S+ filter is helpful to reduce dynamic range loss.
This enables measurements in the 50mV/div to 100 mV/div range which represents a preferred operating point for our oscilloscope.

\section{RP2350 Implementation Concept}\label{sec:impldetails}

Their hardening reports omits any reference to an academic publication.
We hypothesize the implementation might have been derived from~\cite{RDP08}, presenting a second-order secure AES implementation with three full shares.
However, the RP2350hc2 authors apparently chose~\cite{BGG+15} the following simplifications:

\begin{itemize}
\item Instead of 3 full shares for the plaintext or AES state, the third share is common across all of the AES state. They call the resulting implementation a 2.5 sharing where the state $X$ is shared as $X=A \oplus B \oplus C_{\mathrm{state}}$.
\item Instead of 3 full shares for the key, the key uses a third share that is common per word of the AES state, such that the key $K$ is shared as $K=Y \oplus Z \oplus C_{k,\mathrm{word}}$.
\end{itemize}

Due to this not being a full three-share scheme, it is evident that this cannot be a 2nd-order secure scheme anymore.
Reasons for that include, e.g., mask non-uniformity as the 3rd share cannot be chosen independently of other values.
In addition, the hardening report suggest a value probing model was considered.
To prevent transitional leakage, barriers must be used to separate the values.
Especially in a power analysis scenario, it is likely for leakage to still spread unintentionally across barriers due to overlap of leakage points, where the random barrier load `only' adds noise while not fundamentally breaking the relation.

From a side-channel point of view, the implementation uses the following countermeasures: \emph{i}) 2.5 sharing-based table masking, \emph{ii}) barriers with random loads to prevent accidental unsharing, \emph{iii}) a \texttt{makeperm16} called shuffle of the order in which Sbox operations are performed, \emph{iv}) an in-state carry-forward re-randomization of the shares, and \emph{v}) 32-bit register operations wherever possible, except the Sbox processing.

\section{Side-Channel Analysis}\label{sec:sca_analysis}

Our analysis is based on a typical workflow comprising:

\begin{itemize}
\item Visual inspection and leakage detection, see Section~\ref{sec:visleak}
\item Plaintext dependence analysis, see Section~\ref{sec:keydependence}
\item Key extraction, see Section~\ref{sec:keyextraction}
\end{itemize}

Based on the implementation concept, we are most interested in exploring it with a univariate analysis scenario, i.e., to exploit imperfections in the distributions.
It is a well-known mechanism that masked implementations exhibit such imperfections based on the order of their masking/sharing.

\begin{figure*}[!htbp]
\centering
\includegraphics[width=0.95\linewidth]{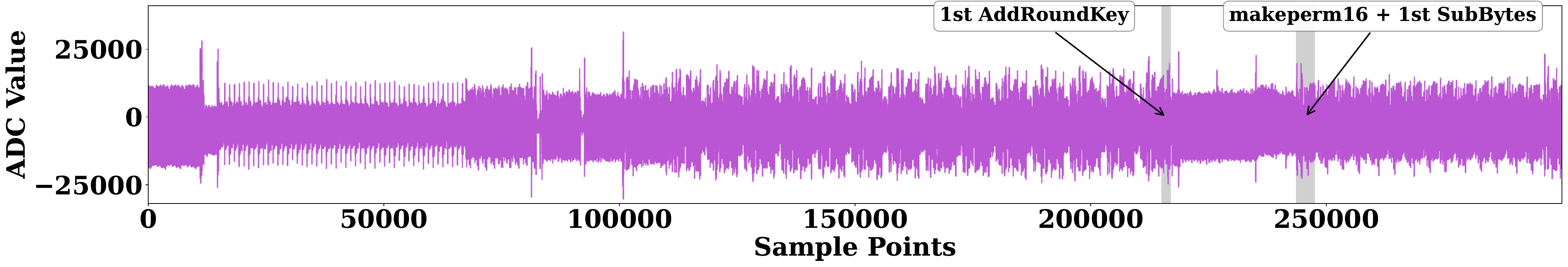}
\caption{Example trace (300 kpts) of the RP2350 AES, showing all relevant activity prior to the falling trigger edge.}
\label{fig:trace_overview}
\end{figure*}

\subsection{Visual Inspection and Leakage Detection}\label{sec:visleak}

A recorded trace is illustrated in Figure~\ref{fig:trace_overview}.
The final section depicts 14 distinct patterns, representative of AES rounds.
The initial key addition was identified through plaintext dependence analysis (cf. Section~\ref{sec:keydependence}).
Our next step is leakage detection for which we recorded 1.5M traces for each set (3M total)~\cite{GJJR11}.

\paragraph{Moments-oriented}

Leakage detection based on Welch's t-test studies differences in statistical moments~\cite{SM15} to assess whether two sets of data are significantly different from each other.
Accordingly, two sets of data are recorded, in our case, a standard fixed vs. random data set~\cite{GJJR11}.
This test is performed by evaluating the $t$-statistic to determine if the null hypothesis can be rejected, i.e., the device is confirmed to leak.
For the sake of simplicity, usually a threshold $|t|>4.5$ is defined to make that determination.

\begin{figure}[htbp]
\centering
\includegraphics[width=0.95\linewidth]{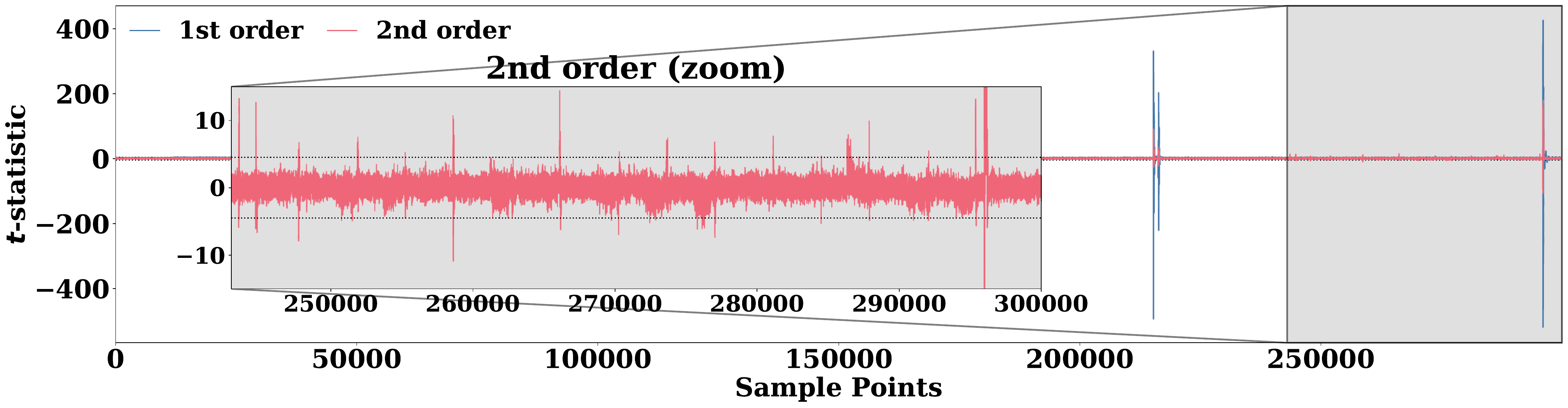}
\caption{Welch's t-test (t-statistic).}
\label{fig:welcht}
\end{figure}

\begin{figure}[htbp]
\centering
\includegraphics[width=0.95\linewidth]{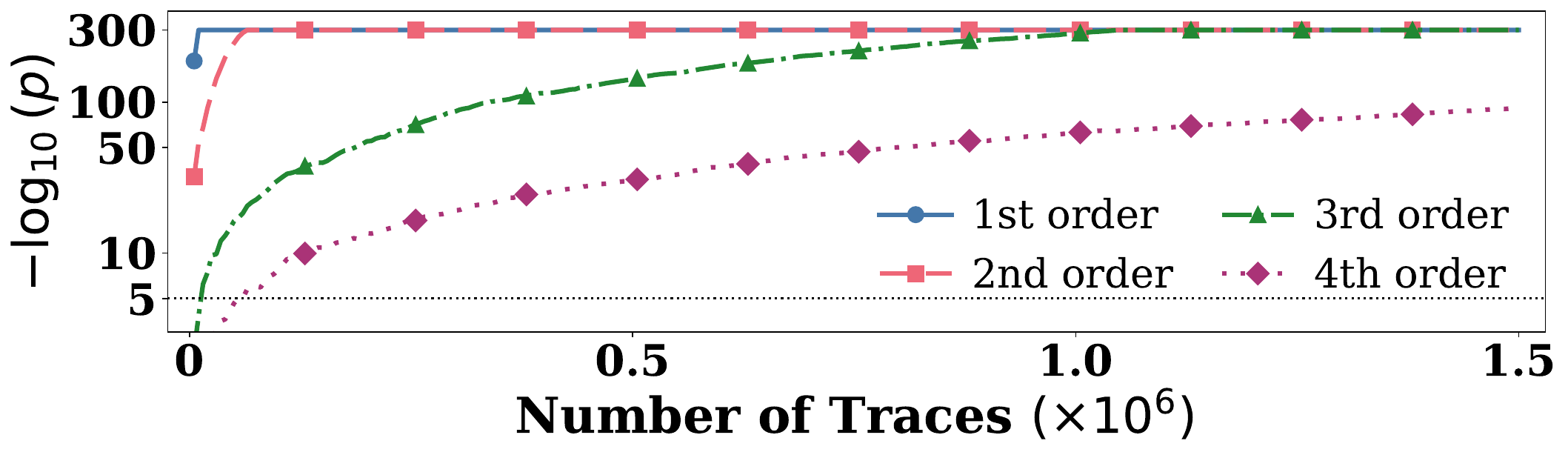}
\caption{MTD plot for Welch's t-test up to 4th order.}
\label{fig:welchtmtd}
\end{figure}

As plotted in Figure~\ref{fig:welcht}, first-order leakage is immediately visible.
Second-order moment leakage emerges for the area of \texttt{makeperm16} and the 14 AES rounds.
The permutation makes this a weak(er) signal but does not prevent discovery.
First-order leakage should be considered unacceptable given the sophistication of the design.
Second-order leakage is likely the result of non-uniformity because of this not being a full three share scheme.
Figure~\ref{fig:welchtmtd} shows the complementary Welch's t-test as a Measurement To Disclosure (MTD) plot.

\paragraph{Deep-Learning}

\begin{figure}[htbp]
\centering
\includegraphics[width=1\linewidth]{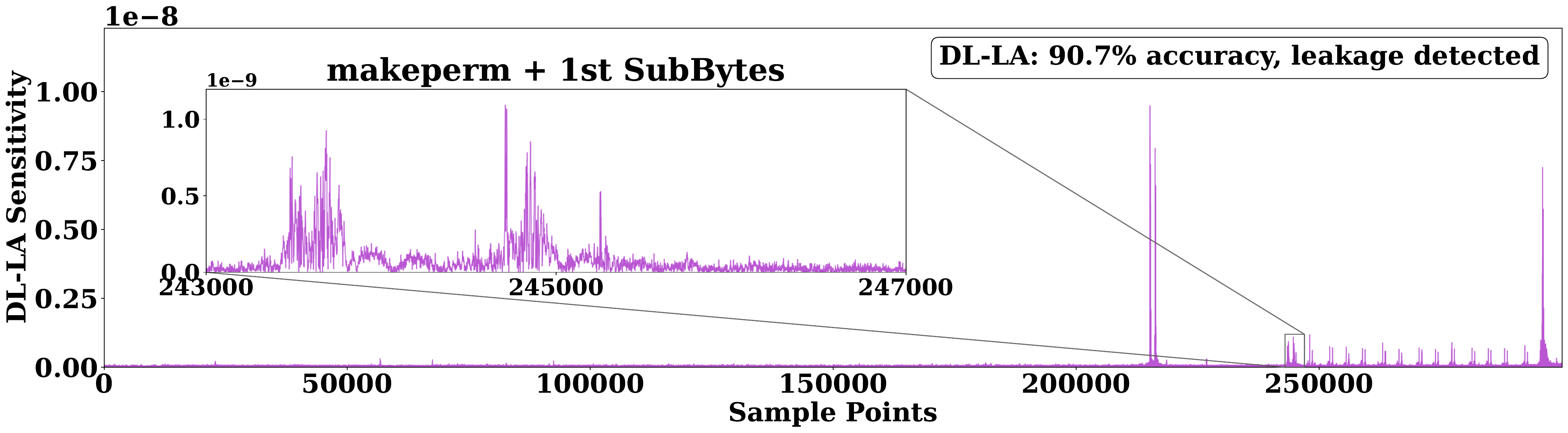}
\caption{Deep-Learning Leakage Asssessment.}
\label{fig:leakage_dlla}
\end{figure}

Leakage-detection can also be evaluated based on deep-learning.
Rooted in the same principles and same data set, we performed Deep-Learning Leakage Assessment (DL-LA)~\cite{MWM21}.
This results in the plot as shown in Figure~\ref{fig:leakage_dlla}, obtained after only 5 epochs.
Clearly visible are all relevant functions of the AES: first key addition, \texttt{makeperm16}, and all rounds of the AES.
It is apparent that the set of countermeasures is insufficient to evade leakage detection.

\subsection{Plaintext Dependence Analysis}\label{sec:keydependence}

\begin{figure}[htbp]
\centering
\includegraphics[width=0.95\linewidth]{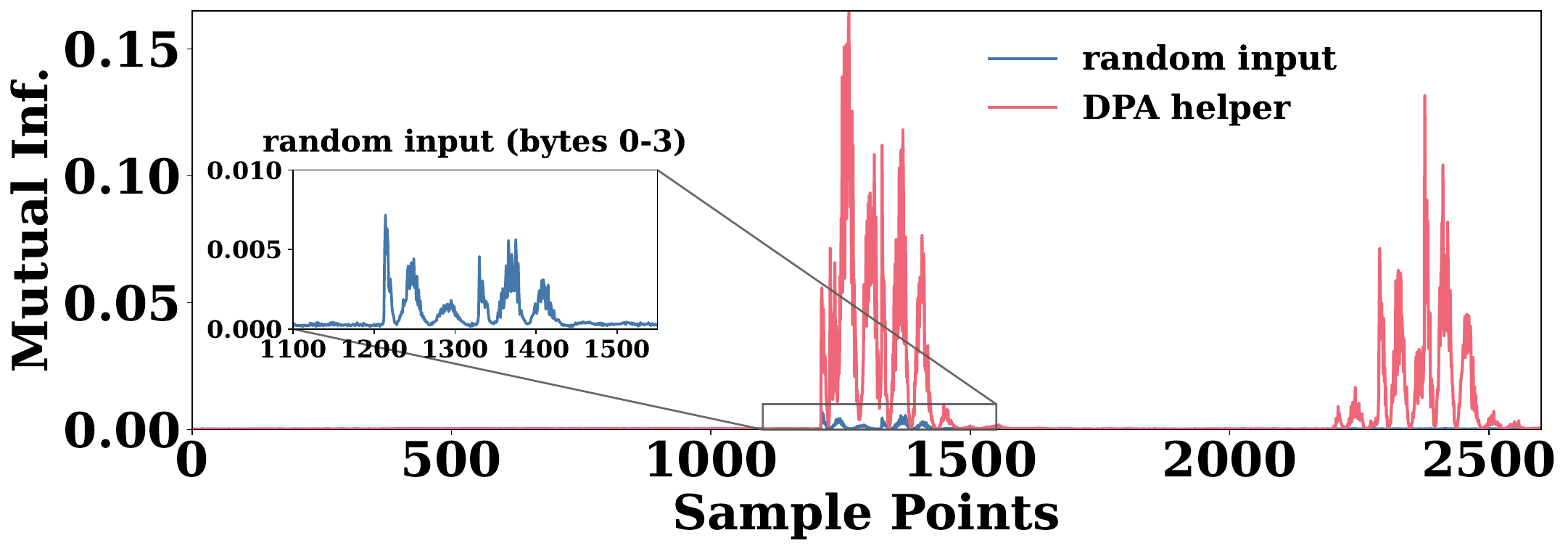}
\caption{Mutual Information-based plaintext dependence.}
\label{fig:pdependence}
\end{figure}

We performed this analysis by computing $\mathrm{I(P,T)}$, with $\mathrm{I}(\cdot)$ being the Mutual Information, $P$ the plaintexts, and $T$ the traces.
This reveals the location of the initial key addition, as shown in Figure~\ref{fig:pdependence}.
Suitable DPA helper patterns can dramatically amplify the dependence, e.g., by only stimulating a single byte position, or by stimulating them in a synchronized way instead of randomizing them all.
If randomized, the mutual information is very faint (0.007).
Since the order for key addition is not permuted, the bytes leak in groups of 4 in ascending order.

With DPA-helper inputs, also co-dependencies across the 4 groups can be revealed, e.g., Figure~\ref{fig:pdependence} is showing plaintext dependence of byte0 in the traces in two different locations.
We consider this undesirable and indicative of implementation flaws.
The first activity regions correspond to bytes0-3 whereas the second activity region responds to bytes4-7.

\subsection{Key Extraction}\label{sec:keyextraction}

For the subsequent work, we recorded different data sets with a random key that is the result of cryptographically hashing (SHA256, 64 rounds) the string ``\emph{rp2350\_hacking\_challenge\_2}'': \texttt{9322982ada8478e0f14d4556a030b944}\\\texttt{a38f34122452d6808022f516ecb539b7}\\
Since we control the cipher input/output, and the challenge did not place restrictions on which round to attack, we focused on the start of the cipher.
A successful attack on the first round will recover the first half of the AES256 key; an attack on the second round the other half.
Trace numbers and data stimuli pattern will be mentioned for each attack.

\begin{figure}[!htbp]
    \centering
    \includegraphics[width=0.9\columnwidth]{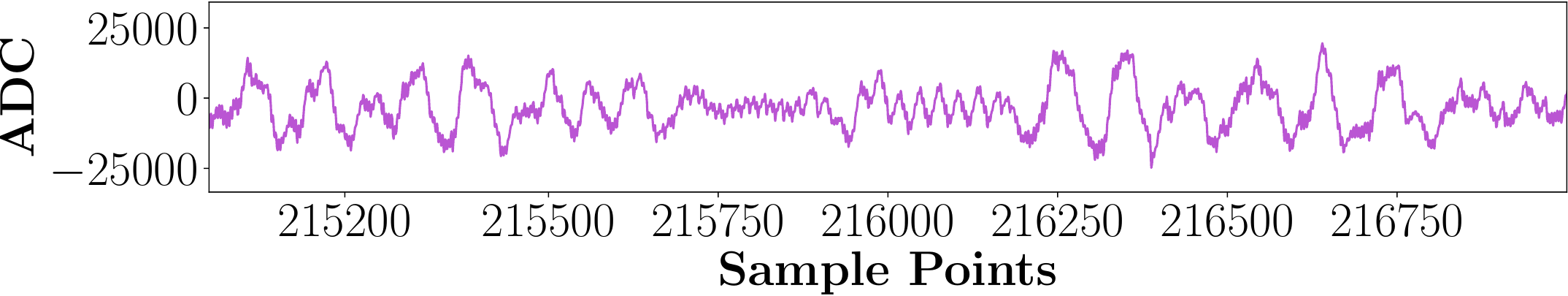}
    \caption{Relevant range of first key addition.}
    \label{fig:trace_zoom_key_add}
\end{figure}

\paragraph{Sanity Check} By using a DPA-helper input scheme, we discovered a trivial recovery of byte0 on the key addition range illustrated in Figure~\ref{fig:trace_zoom_key_add}.
This particular byte appears to be susceptible to a standard Correlation Power Analysis with Hamming Distance (HD) model which aligns with the assumed susceptibility towards a lower-order transitional leakage model.

\begin{figure}[htbp]
\centering
\includegraphics[width=0.9\linewidth]{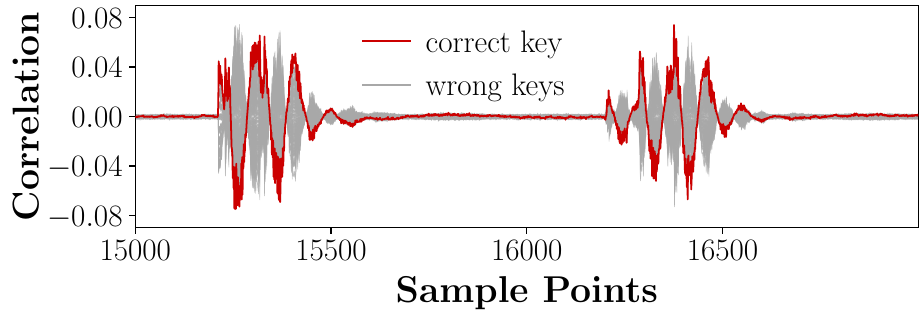}
\caption{Correlation Power Analysis of byte0 with HD model.}
\label{fig:cpatime}
\end{figure}

\begin{figure}[htbp]
\centering
\includegraphics[width=0.9\linewidth]{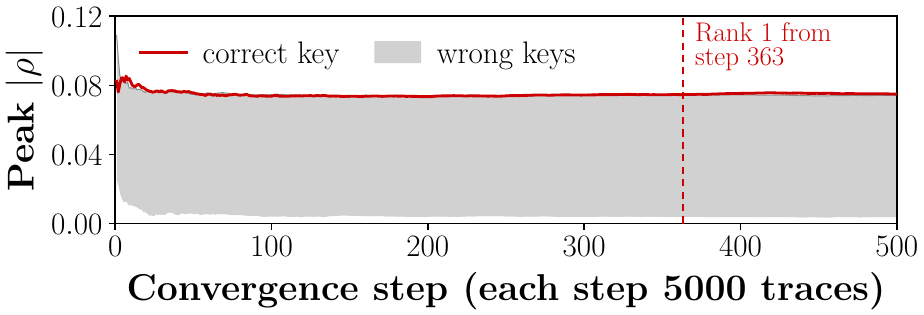}
\caption{MTD plot for CPA of byte0.}
\label{fig:cpamtd}
\end{figure}

While the recovery is narrow (cf. Figures~\ref{fig:cpatime} and~\ref{fig:cpamtd}), and should be considered incidental as it appears it cannot be extended to other bytes, it clearly shows the limits of the current protection of the key addition.
Furthermore, as can be seen in the results, a ``lift-off'' for \emph{all} key candidates can be observed that we consider atypical.
This is due to the substantial plaintext leakage which is (randomly) tied to different key candidates.

\paragraph{SCATTER}

Since key addition appears to be less protected (possibly as a result of a flaw in the implementation) and will likely rely on DPA-helper inputs for successful attacks, we focused our efforts on the seemingly better protected Sbox implementation using SCATTER~\cite{VTM21}.
We consider this the most generic attack that is independent of input patterns and the round being targeted, i.e., successfully defeating the Sbox protections constitutes a full break of the implementation.

\begin{figure}[!thbp]
\centering
\includegraphics[width=0.95\linewidth]{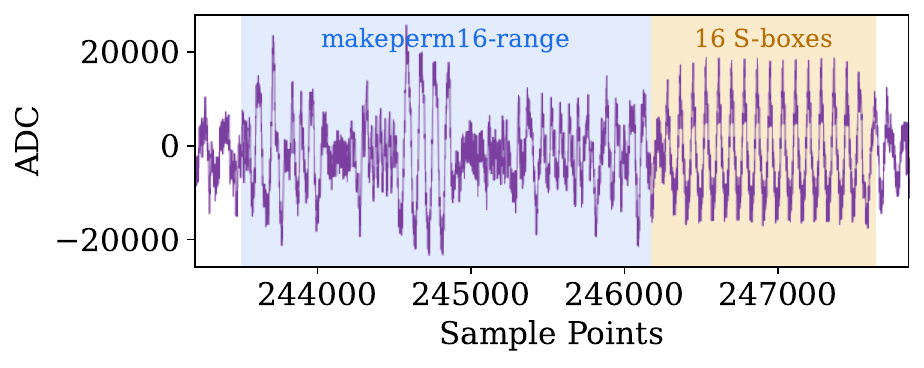}
\caption{SCATTER with \texttt{makeperm16} search range.}
\label{fig:scattertime}
\end{figure}

The Sbox creates a distinct pattern that, combined with insight from second-order moment and DL-LA leakage, can be conveniently located. 
Consequently, for carrying out the attack, we selected the whole Sbox range and used it as a fixed reference.
We then ``scatter-slide'' along the suspected \texttt{makeperm16} range with different parameters to combine this into a higher-order side-channel mixture (cf. Figure~\ref{fig:scattertime}).
This results in a successful key recovery of \emph{every} byte (rank 1) with a median MTD range of approx. 1.1M for all key bytes, nicely confirming our leakage detection and extraction efforts.
Both HD and HW models are applicable to the mixture.

\begin{figure}[!thbp]
\centering
\includegraphics[width=0.95\linewidth]{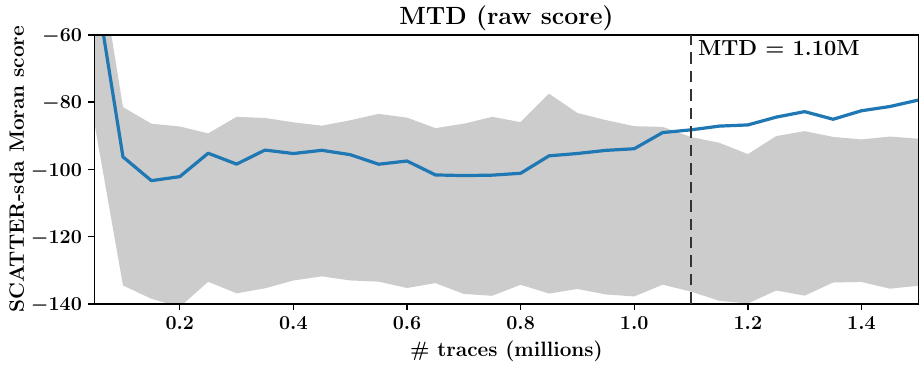}
\caption{MTD SCATTER plot for representative byte.}
\label{fig:scattermtd}
\end{figure}

Since we do not exclude portions where barrier loads happen, attack performance is limited by their added noise.
A more tailored approach is likely to substantially improve over our results, however, within a black-box scenario, we consider the key extraction to be fully completed with a reasonable amount of traces that is of practical relevance. 
The exact parameters that we used for our attack are: HW model, 80 sample window for \texttt{makeperm16} at approx. 246000, 1300 sample window for Sbox starting at approx. 246200, 64 histogram bins, and a Moran configuration of (5x7x7)=(class,bin,bin).

\section{Conclusion}\label{sec:concl}

We presented a successful side-channel analysis of the RP2350 hacking challenge 2.
This is a limited assessment of basic side-channel properties and not a substitute for a thorough root cause analysis.
We merely exploited what we consider as the most convenient way to demonstrate some of the relevant vulnerabilities of the implementation.

Given the tooling and expertise needed to perform this attack, it is fair to say that the implementation resists standard univariate attacks well despite failing basic leakage detection tests.
Also, as stated in the RP2350 hardware design guide regarding the DVDD supply, ``\emph{A great deal of work [\dots], with many iterations of the PCB required [\dots]}'', likewise, we needed many iterations to obtain high-quality measurements.
However, we are convinced other groups could attack this, too.
Especially if a reward in due proportion to the effort had been offered.

Major parts of our work were performed with SCARR~\cite{BEBI24} but also Claude AI.
AI for SCA greatly amplifies an analyst's capabilities.
Luckily, initial engineering, critical insight, and direction at least for now, needed to be fully human-driven.

More leakage points were identified with a credible risk of a workable extraction method.
We encourage the challenge authors to revise the core construction of their implementation, as the presented attack will likely scale to their full implementation with somewhat limited additional pre-processing.
We leave this improvement and analysis as future work.

\clearpage
\printbibliography

\end{document}